\documentclass[twocolumn,pra,aps,superscriptaddress]{revtex4}

\usepackage{color}
\usepackage{epsfig}
\usepackage{latexsym}
\usepackage{amssymb}
\usepackage{amsmath}
\usepackage{algorithm}
\usepackage{algorithmic}
\usepackage{wrapfig}
\usepackage[apple mac]{inputenc}
\usepackage[english]{babel}
\usepackage{times}
\usepackage{latexsym}
\usepackage{fancyhdr}
\usepackage{verbatim}
\usepackage{tabularx}
\usepackage{epsfig}
\usepackage{amsmath}
\usepackage{amssymb}
\usepackage{graphicx}
\usepackage{wasysym}
\usepackage{hyperref}
\usepackage{tcolorbox}
\usepackage{mdframed}
\usepackage{lipsum,framed}
\usepackage{yfonts}

\newcommand{\qed}{\hspace*{\fill}$\square$}

 \newtheorem{thm}{Theorem}
 \newtheorem{lem}{Lemma}
 \newtheorem{cor}{Corolary}
 \newtheorem{defn}[thm]{Definition}
 
 \newenvironment{proof}{\noindent \emph{Proof.}}{\qed}

\newcommand{\be}{\begin{equation}}
\newcommand{\ee}{\end{equation}}

 \newcommand{\ket}[1]{|#1\rangle}
 \newcommand{\bra}[1]{\langle #1|}

\begin{document}

\title{Homomorphic Encryption of the k=2 Bernstein-Vazirani Algorithm}
\author{Pablo Fern\'andez} 
\affiliation{Departamento de F\'{\i}sica Te\'orica, Universidad Complutense, 28040 Madrid, Spain.}
\author{Miguel A. Martin-Delgado}
\affiliation{Departamento de F\'{\i}sica Te\'orica, Universidad Complutense, 28040 Madrid, Spain.}
\affiliation{CCS-Center for Computational Simulation, Campus de Montegancedo UPM, 28660 Boadilla del Monte, Madrid, Spain.}

\begin{abstract} 
The nonrecursive Bernstein-Vazirani algorithm was the first quantum algorithm to show a superpolynomial improvement over the corresponding best classical algorithm. Here we define a class of circuits that solve a particular case of this problem for second-level recursion. This class of circuits simplifies the number of gates $T$ required to construct the oracle by making it grow linearly with the number of qubits in the problem. We find an application of this scheme to quantum homomorphic encryption (QHE) which is an important cryptographic technology useful for delegated quantum computation. It allows a remote server to perform quantum computations on encrypted quantum data, so that the server cannot know anything about the client's data. Liang developed QHE schemes with perfect security, $\mathcal{F}$-homomorphism, no interaction between server and client, and quasi-compactness bounded by $O(M)$ where M is the number of gates $T$ in the circuit. Precisely these schemes are suitable for circuits with a polynomial number of gates $T/T^{\dagger}$. Following these schemes, the simplified circuits we have constructed can be evaluated homomorphically in an efficient way.
		
\end{abstract}

\maketitle

\section{Introduction}
\label{sec:intro}

Quantum computing has gained great interest in recent decades due to the potential advantages it could offer compared to classical computing. Using quantum phenomena such as superposition and entanglement, quantum algorithms, including hybrid ones, are being sought that are faster than their corresponding better known classical version. Initial examples of such algorithms were Grover's search algorithm \cite{Grover} or Shor's factorization algorithm \cite{Shor}. The search continues as part of developments in current quantum technologies.

The first hint of the possible advantage of quantum computers came from Deutsch \cite{Deusch1} and Deutsch-Jozsa \cite{Deuts}. In the latter work a quantum computer could be used deterministically to solve a problem in one go. This problem would require an exponential number of queries on a classical computer using an exact model where no probability of failure was allowed. However, when a bounded error was allowed in the algorithm, this advantage disappears.

Soon after, Bernstein and Vazirani
introduced a problem in which the query complexity of the quantum solution was better than what was possible by any classical computer. The Bernstein-Vazirani (BV) \cite{BV} algorithm is the first quantum algorithm that showed the advantages that a quantum computer could have over a classical one. The nonrecursive version of the problem improved the query complexity from $\Omega(n)$, which was the best a classical computer could achieve, to $\Omega(1)$ for the quantum computer. This constitutes a polynomial speed-up. This algorithm has been used as the basis of different applications. As an example of an application in cryptography, the BV algorithm has been used to find the linear structures of a function in order to attack block ciphers \cite{Block ciphers}. A variant of
the algorithm was studied by Cross et al. \cite{apli1} for quantum learning robust against noise.

The nonrecursive BV algorithm is classically tractable because the hardness of the classical problem scales only polynomial in the problem size. Bernstein and Vazirani managed to find a version of the problem which is classically intractable, but can be solved on a quantum computer using polynomial resources. This problem is known as the recursive Bernstein-Vazirani problem. This problem has an exponential query complexity in the classical realm and only a polynomial one in the corresponding quantum version. Therefore the recursive Bernstein-Vazirani algorithm has a super polynomial speed-up compared to the best classical algorithms \cite{Bacon}. This problem is one of the two building blocks of this paper. The other is quantum homomorphic encryption.

In the current era of cloud computing, enormous quantities of private data have been uploaded to the cloud by users for storage and many other services. Computations performed on the cloud should be encrypted to preserve the security of the data, just like modern cryptography is used to ensure the security of communications. Homomorphic encryption allows a client to encrypt his data, send it to a server that will operate on this encrypted data and then returns the data once all the operations have finished so the decryption of data can be done. This way the server can never learn anything about the actual data it is operating with. Since the first classical fully homomorphic encryption scheme (FHE) created by Gentry \cite{Gentry}, many other classical schemes have been proposed and perfected. The security of these schemes is based on the difficulty of certain mathematical problems, such as ideal lattices \cite{Gentry} or the learning with errors problem (LWE) \cite{LWE} .

In terms of quantum computing, we have now seen the development of real, incipient quantum computers. Some of them can be accessed in the cloud, such as the IBM Q quantum computers, allowing researchers to perform different types of protocols and experiments on them. Since most quantum computers in the foreseeable future will be accessible through the cloud, quantum homomorphic encryption (QHE) schemes have been developed to ensure their security. QHE allows a remote server to perform some quantum circuit $QC$ on encrypted quantum data $Enc(\rho)$ provided by a client, and then the client can decrypt the
server's output and obtain the result $QC(\rho)$. The security of these schemes depend on the fundamental properties of quantum mechanics instead of the difficulty of mathematical problems.

These schemes can be constructed with interaction between client and server or without it (interaction in this context means that the client and server communicate during the execution of the circuit). Constructing a scheme with interactions is easier than constructing one without them. Allowing interactions decreases the efficiency of the scheme because the server will have to wait for the client to perform some operations before it can continue with the scheme it was executing. Some Quantum Fully Homomorphic Encryption (QFHE) schemes where interaction is allowed have been constructed, like Liang's \cite{T interactions} scheme, where the total amount of interactions is the same as the number of $T$ gates in the circuit. However, we are more interested in schemes without interaction between client and server. 

The research of QHE schemes began in 2012. Rohde \cite{rhode} proposed a symmetric-key QHE scheme, which allows quantum random walk on encrypted quantum data. This protocol has been realized recently by Zeuner et al. in \cite{experimento}. This scheme is not $\mathcal{F}$-homomorphic, which means it can not perform universal quantum computation on homomorphic encrypted data (it only allows the homomorphic evaluation of certain circuits).

Tan \cite{Tan} proposed a QHE scheme which can
perform extensive quantum computation on encrypted data, but it can not provide security in a cryptographic sense since it can only hide $n/ \log n$ bits, where $n$ is the total size of the message, so the ratio of the hidden amount of information to the total amount approaches 0 when $n$ approaches infinity.

Yu et al. proved a no-go result in \cite{no go result}, which states that any Quantum Fully Homomorphic Encryption scheme (QFHE) with perfect security must produce exponential storage overhead if interaction between client and server is not allowed. This means that it is impossible to construct an efficient QFHE scheme with perfect security and no interaction.

Lai and Chung \cite{enhanced no go} proved an enhanced no-go result, which states that it is impossible to construct a non-interactive and information theoretically secure (ITS) QFHE scheme. This means that if non-interaction is required, the best security a QFHE scheme could hope to achieve is computational security (security based on hard mathematical problems just like classical cryptography). Lai
and Chung proceeded to construct a non interactive ITS QHE scheme with compactness, which means that the complexity of the decryption procedure does not depend on the operations of the evaluated quantum circuit. However, due to the no-go result, it is not $\mathcal{F}$-homomorphic, so it is just partially homomorphic. In particular it only allows the homomorphic evaluation of a class of quantum
circuits called instantaneous quantum polynomial-time (IQP) circuits. Since it is not fully homomorphic, it is a Quantum Somewhat Homomorphic Encryption scheme (QSHE).

Attempts have been made to study possible schemes with less stringent conditions. The relevant properties of QHE are non-interaction, $\mathcal{F}$-homomorphism, compactness and perfect security. Due to the no-go result explained above, it is impossible to have all properties at the same time. If interaction is allowed, it is possible to construct a QHE with perfect security, such as the one already mentioned by Liang \cite{T interactions}. Broadbent and Jeffery \cite{Broadbent} constructed two non-interactive quantum homomorphic schemes, combining quantum one-time pad (QOTP) and classical fully homomorphic encryption (FHE), so they downgrade from perfect security to security bounded by classical homomorphic encryption schemes. The complexity of the  decryption procedure of the first scheme scales with the square of the number of $T$-gates. The second scheme utilizes a quantum evaluation key, whose length is given by a polynomial of degree exponential in the circuit's $T$-gate depth, in order to generate a homomorphic scheme for quantum circuits with constant $T$-depth. 

Liang \cite{Liang} constructed two non-interactive and perfectly secure QHE schemes, which are $\mathcal{F}$-homomorphic but not compact. In fact, these schemes are quasi-compact (the decryption procedure has complexity that scales sublinearly
in the size of evaluated circuit) as defined by Broadbent and Jeffery in \cite{Broadbent}. The first scheme in \cite{Broadbent} is known as EPR, with quasi-compactness, $\mathcal{F}$-homomorphism, non-interaction and computational security. EPR was proved to be $M^2$-quasi-compact, where $M$ is the number of $T$-gates in an evaluated circuit. Both Broadbent's and Liang's schemes makes use of Bell
states and quantum measurements. Liang's schemes \cite{Liang} are better regarding security, because they are perfectly secure whereas EPR is only computationally secure. Also one of Liang's schemes, VGT, is $M$-quasi-compact (the complexity of the  decryption procedure scales with the number of $T$-gates), so it is also better in that regard. As these schemes are quasi-compact, they do not contradict the no go-result.

The importance of Liang's schemes \cite{Liang} is that even though they are quasi-compact, they allow the homomorphic evaluation of any quantum circuit with low $T/T^\dagger$ gate complexity with perfect security. The decryption procedure would be inefficient for circuits containing an exponential number of $T/T^\dagger$ gates, but it is suitable for circuits with polynomial number of $T/T^{\dagger}$gates. Liang's schemes have been used to implement a cyphertext retrieval scheme based on the Grover algorithm in Gong et al. \cite{QHE  grover}.

The paper is organized as follows. In section \ref{sec:review} the nonrecursive and the recursive version of the BV algorithm are reviewed. In section \ref{sec:definition} the definition of a certain type of circuits that solve the recursive BV problem for $k=2$ with certain important properties will be given and characterized. In section \ref{sec:application} Quantum Homomorphic Encryption is reviewed and the homomorphic implementation of the recursive BV problem is discussed. Finally, the conclusions are summarized in section \ref{sec:conclusions}.

\section{Review of the Bernstein-Vazirani algorithm}
\label{sec:review}
In the BV problem a $n$ bit function $f: \{0,1\}^n \rightarrow \{0,1\}$ is given. The function obeys the relation: $f_s(x)= s \cdot x=x_1s_1+ x_2s_2+....+x_ns_n \mod(2)$ where $s$ is an unknown $n$ bit string and the objective is to find $s$. The best classical algorithm to solve this problem in the exact query complexity model takes $n$ queries since the function only returns one bit, so the best algorithm is simply using as input a bit string where all bits are a 0 except the $i$th bit which would be a 1 in order to obtain $s_i$ from the function. If bounded error is allowed, suppose the function could be queried in a probabilistic way so all of the bits of $s$ could be found in less than $n$ queries
with some probability of failure that is bounded below 1/2. This bounded error algorithm would still need $\Omega(n)$ queries \cite{Bacon}. This contrasts with the Deutsch-Jozsa algorithm because a classical probabilistic algorithm makes the quantum advantage disappear.

Regarding the quantum algorithm, the main idea is implementing $f_s(x)$ in a reversible way using a quantum oracle:
\begin{equation}
U_s= \sum_{x\in \{0,1\}^n} \sum_{y \in \{0,1\}} \ket{x}\bra{x} \otimes \ket{y \oplus (s \cdot x)}\bra{y}.
\label{oracle1}
\end{equation}
The quantum algorithm uses $n$ qubits all starting in the $\ket{0}$ state. It begins by applying $n$ $H$ gates, one to each qubit. It will result in an equal superposition of all possible $n$-bit strings:
\begin{equation}
H^{\otimes n}\ket{00...0}=\frac{1}{\sqrt{2^n}}\sum_{x\in\{0,1\}^n}\ket{x}.
\end{equation}
This will be the input for the oracle. The oracle will be implemented using an extra qubit in the state $\ket{-}=\frac{\ket{0}-\ket{1}}{\sqrt{2}}$. If a $CNOT$ is applied to the $\ket{-}$ state as the target qubit, a phase kickback will occur where the control qubit will change its sign and the target qubit will remain unchanged. This is because if a $CNOT$ is applied to the $\ket{-}$ state with $\ket{1}$ as the control qubit: $CNOT[\ket{1}\otimes\frac{\ket{0}-\ket{1}}{\sqrt{2}}]=\ket{1}\otimes\frac{\ket{1}-\ket{0}}{\sqrt{2}}=\ket{1}\otimes(-\frac{\ket{0}-\ket{1}}{\sqrt{2}})=-\ket{1}\otimes\frac{\ket{0}-\ket{1}}{\sqrt{2}}$. This means the phase of the terms where $s \cdot x =1$ will be flipped so after applying $U_s$:
\begin{equation}
\ket{\phi}_s=\frac{1}{\sqrt{2^n}}\sum_{x\in\{0,1\}^n}(-1)^{f_s(x)}\ket{x}=\frac{1}{\sqrt{2^n}}\sum_{x\in\{0,1\}^n}(-1)^{s\cdot x}\ket{x}.
\end{equation}
Since the action of the Hadamard gate on a general computational basis state of a $n$-qubit system is given by:
\begin{equation}
 H^{\otimes n}\ket{u}=\frac{1}{\sqrt{2^n}}\sum_{x\in\{0,1\}^n}(-1)^{u\cdot x}\ket{x}.
 \label{h1}
\end{equation}
and the $H$ gate is its own inverse:
\begin{equation}
H^{\otimes n} \left[\frac{1}{\sqrt{2^n}}\sum_{x\in\{0,1\}^n}(-1)^{u\cdot x}\ket{x}\right] =\ket{u}.
\label{h2}
\end{equation}

\begin{figure}[]
	\includegraphics[width=0.3\textwidth]{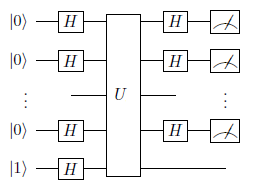}
	\centering
	\caption{nonrecursive Bernstein-Vazirani circuit. 
	}
	\label{Us}
\end{figure}

Then if we now apply again $n$ $H$ gates, according to equation \eqref{h2} the state of the qubits will be $\ket{s}$:
\begin{equation}
H^{\otimes n}\ket{\phi}_s=H^{\otimes n}\left[\frac{1}{\sqrt{2^n}}\sum_{x\in\{0,1\}^n}(-1)^{s\cdot x}\ket{x}\right]=\ket{s}
\end{equation}
Finally once the qubits are measured $s$ will be obtained with a 100\% probability. The algorithm finds the value of $s$ in just one query. The circuit that solves the BV problem is represented in figure \ref{Us}.

In the nonrecursive BV problem, we are given a function $f_s(x)=s\cdot x$ and $s$ has to be found. In the recursive BV problem, what has to be found is some function $g: \{0,1\}^n \rightarrow 	\{0,1\}$ on $s$, $g(s)$. We will refer to all the possible $n$ bits strings that can be used as inputs to the function $g$ by $s_{in}$ $\in  \{0,1\}^n$. 

To construct the first level of the recursion, an oracle based on two $n$ bits strings $x\in \{0,1\}^n$ and $y\in \{0,1\}^n$ is used. As in the nonrecursive BV algorithm the function queried takes these two bit strings as input and outputs $s_x\cdot y$, so the function can be expressed as $f:\{0,1\}^n \times \{0,1\}^n \rightarrow \{0,1\}$ given by $f(x,y)=s_x\cdot y$. The $s_x$ are $2^n$ different bit strings labelled by $x\in\{0,1\}^n$ with one important property: the value of $g(s_x)$ has to coincide with the product of $s \cdot x$ for some unknown $s$. The two level BV problem then is to identify this $s$ and $g(s)$.

When $f(x,y)$ is queried, $s_x \cdot y$ is obtained. To identify $s_x$ for a fixed $x$, the original nonrecursive problem would have to be solved. Then once $s_x$ has been obtained for a fixed $x$, $s_x$ has to be used to calculate $g(s_x)$. However each  $g(s_x)$ for different values of $x$ is now part of a nonrecursive BV problem, so this way the problem becomes more complex.

This process of creating the two level BV problem can be done $k$ times to create the $k$ level BV problem. At the $k$th level the function $f$ takes as input $k$ strings of $n$ bits denoted by $x_i$ and computes $f(x_1,x_2,....,x_k)=x_k\cdot s_{x_1,x_2,...x_{k-1}}$. The secret bit string $s_{x_1,x_2,...x_{k-1}} \in \{0,1\}^n$ produces the next lower level problem since $g(s_{x_1,x_2,...x_{k-1}})= x_{k-1}\cdot s_{x_1,x_2,...x_{k-2}}$ where as before $s_{x_1,x_2,...x_{k-2}}\in \{0,1\}^n$. This process can be continued until the final level is reached where $g(s_{x_1})=s\cdot x_1$. At this level $s$ can finally be found.

We will be discussing the recursive Bernstein-Vazirani problem for $k=2$ from this point onwards.

We will first solve the problem for an arbitrary case in $k=2$. As in the usual BV problem, we are given access to a unitary which computes in a reversible manner $f(x_1,x_2)=s_{x_1} \cdot x_2$. Using the usual BV algorithm the value of a $s_{x_1}$ for a fixed $x_1$ would be easy to find. The tricky part of the recursive BV problem is that when $g$ is applied to these $s_{x_1}$s we get a new BV problem to solve.
This implies that we have to use the BV algorithm over all $x_1$s as well.

The oracle for the two level problem is:
\begin{multline}
U_s= \sum_{x_1, x_2 \in \{0,1\}^n} \sum_{y \in \{0,1\}} \ket{x_1}_1\bra{x_1}_1 \otimes \\ \ket{x_2}_2\bra{x_2}_2 \otimes \ket{y \oplus (s_{x_1} \cdot x_2)}\bra{y}
\end{multline}

If we start with two quantum registers, $x_1$ and $x_2$ with all their qubits in the $\ket{0}$ state and apply $H$ gates to both of them we will get the usual state containing the superposition over all possible bitstrings on the $x_1$ and $x_2$ registers:

\begin{equation}
H^{\otimes 2n}\ket{00...0}=
\frac{1}{2^n} \sum_{x_1, x_2 \in \{0,1\}^n} \ket{x_1}_1 \otimes \ket{x_2}_2.\label{superposition}
\end{equation}

The two registers are labelled by a $1$ and $2$ subscript respectively.
If we use the usual phase-back trick using a qubit in the $\ket{-}$ state as the target qubit we can implement the oracle $U_s$ and we will have the following state:

\begin{equation}
\frac{1}{2^n}  \sum_{x_1, x_2 \in \{0,1\}^n} (-1)^{s_{x_1}\cdot x_2} \ket{x_1}_1 \otimes \ket{x_2}_2 \otimes \ket{-}.
\end{equation}

By performing the $n$ qubit Hadamard gate on the second register, we see that it will transform this state to:

\begin{equation}
\frac{1}{\sqrt{2^n}}  \sum_{x_1 \in \{0,1\}^n} \ket{x_1}_1 \otimes \ket{s_{x_1}}_2 \otimes \ket{-}.
\label{eq:10}
\end{equation}

We now must apply $g$ to $s_{x_1}$ to solve the next order Bernstein-Vazirani problem, since $g(s_{x_1}) = s \cdot x_1$. The oracle for $g$ is given by:
\begin{equation}
G=\sum_{x\in \{0,1\}^n} \sum_{y \in \{0,1\}} \ket{x}_2\bra{x}_2 \otimes \ket{y \oplus g(x)}\bra{y}
\label{G}
\end{equation}

If we attach an extra ancilla qubit in the state $\ket{-}$, we can use the phase-back trick to use this unitary to turn eq.\eqref{eq:10} into:

\begin{equation}
\frac{1}{\sqrt{2^n}}  \sum_{x_1 \in \{0,1\}^n} (-1)^{g(s_{x_1})} \ket{x_1}_1 \otimes \ket{s_{x_1}}_2 \otimes \ket{-}  \otimes \ket{-}.
\end{equation}

This oracle acts only on the second register and the second extra qubit in the $\ket{-}$ state.
Next, we need to apply the $n$ qubit Hadamard to the second register and then we apply our oracle $U_s$ again. After applying the $H$ gates:

\begin{equation}
\frac{1}{2^n}  \sum_{x_1, x_2 \in \{0,1\}^n} (-1)^{g(s_{x_1})} (-1)^{s_{x_1}\cdot x_2} \ket{x_1}_1 \otimes \ket{x_2}_2 \otimes \ket{-} \otimes \ket{-}.
\end{equation}
After applying $U_s$ we get another phase kickback:

\begin{multline}
\frac{1}{2^n}  \sum_{x_1, x_2 \in \{0,1\}^n}  (-1)^{g(s_{x_1})} (-1)^{s_{x_1}\cdot x_2} (-1)^{s_{x_1}\cdot x_2}\\ \ket{x_1}_1 \otimes \ket{x_2}_2 \otimes \ket{-} \otimes \ket{-} = \\ \frac{1}{2^n}  \sum_{x_1, x_2 \in \{0,1\}^n}  (-1)^{g(s_{x_1})}  \ket{x_1}_1 \otimes \ket{x_2}_2 \otimes \ket{-} \otimes \ket{-}. 
\end{multline}

Then, we can apply $H$ gates to all the qubits in the first register so we get the following state:
\begin{equation}
\frac{1}{\sqrt{2^n}}  \sum_{x_2 \in \{0,1\}^n}  \ket{s}_1 \otimes \ket{x_2}_2 \otimes \ket{-}  \otimes \ket{-}.
\end{equation}

\begin{figure}[]
	\includegraphics[width=0.4\textwidth]{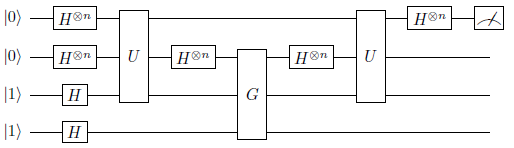}
	\centering
	\caption{Recursive Bernstein-Vazirani circuit for level-2 recursion.}
	\label{instrucciones}
\end{figure}

Finally, by measuring the first register we will get $s$ with 100\% probability. The recursive BV circuit represented graphically is shown in figure \ref{instrucciones}.

\begin{figure}[]
	\includegraphics[width=1.02\columnwidth, height=0.32\linewidth]{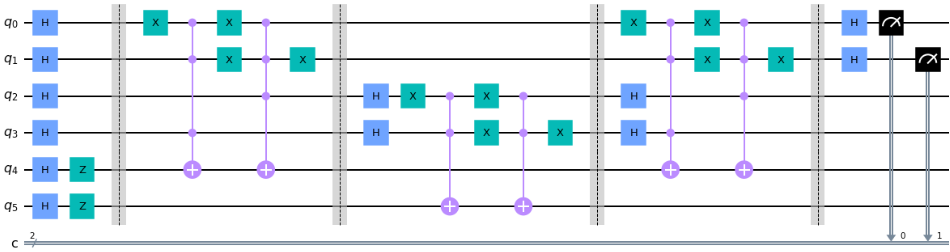}
	\centering
	\caption{Recursive Bernstein-Vazirani circuit for two recursions for $s=11$, represented in Qiskit.}
	\label{mcnots}
\end{figure}

The oracle $U_s$ for this problem in the most general case is made of multi-controlled $CNOT$ gates where $n$ control qubits are used for the first register and one for the second. As an example, consider the recursive BV problem for $k=2$ where the secret bit string is $s=11$ and the values of $g(s_{in})$ are chosen as: $g(00)=0$, $g(01)=g(10)=g(11)=1$. Then the choice of values $s_{00}=s_{11}=00$, $s_{01}=01$, $s_{10}=10$ is compatible with the fact that $g(s_{x_1})=s\cdot x_1$. The circuit that solves this problem is represented graphically in figure \ref{mcnots}, where the oracle $U_s$ is implemented using multi-controlled $CNOT$ gates.

However in certain situations $U_s$ can be implemented just using Toffoli gates. We will define the circuits where this is possible in the following section.

\section{T-lineal circuits with reduced circuit complexity}
\label{sec:definition}

We will begin this section defining \textit{T-lineal circuits with reduced circuit complexity (RCC)}.

\begin{tcolorbox}
	[colback=white]
	\begin{defn}
		\textit{T-lineal circuits with reduced circuit complexity (RCC)}  are quantum circuits that solve the Bernstein-Vazirani recursive problem when the oracle $U_s$ can be constructed just with $n$ Toffoli gates at most and the oracle $G$ just with $CNOT$ gates.
	\end{defn}
\end{tcolorbox}

Here $n$ is the same number as the qubits contained in one of the registers of the algorithm and also the number of bits needed to represent s. We will now discuss for which values of $g(s_{in})$ and $s_{x_1}$ a \textit{T-lineal circuit with reduced circuit complexity (RCC)}  can be constructed, starting with lemma $\ref{def1}$:

\begin{tcolorbox}
	[colback=white]
	\begin{lem}
		Given a function $g:\{0,1\}^n\rightarrow \{0,1\}$ on $s_{in}$, $g(s_{in})$, that also obeys $g(s_{x_1})=s \cdot x_1$. A T-lineal circuit with reduced circuit complexity (RCC) can be constructed in which $g(s_{in})=s \cdot s_{in}$ and $s_{x_{1}}=x_1 \wedge s$  $\forall x_1$ where $x_1 \wedge s$ represents the bitwise AND operation between  $x_1$ and $s$, 
		where $s$ and $x_1$ are strings, $s \in \{0,1\}^n$, $x_1 \in \{0,1\}^n$.
		\label{def1}
	\end{lem}
\end{tcolorbox}

  \begin{proof}
  	We want to prove that the values of $g(s_{in})$ and $s_{x_1}$ from lemma \ref{def1} can be implemented using $CNOT$s and Toffolis respectively.

  	We will start with $g(s_{in})$. In the Bernstein-Vazirani problem any value for $g(s_{in})$ can be chosen as long as it equals 0 and 1 for a least one of its inputs. Also in all Bernstein-Vazirani circuits $g(s_{x_1})=s\cdot x_1$. If the values of $g(s_{in})$ fulfil $g(s_{in})=s \cdot s_{in}$ then the values of $g(s_{in})$ will depend on the value of $s$ in two ways: whether or not a bit in $s$ is a 1 and the position of these 1s. The product $s\cdot s_{in}$ will result in a 1 when $s_{in}$ has an odd number of 1s in the same positions as the 1s in $s$. In other words, for every 1 in $s$ in the i-th position, the i-th qubit of the second register of the circuit will experiment a phase kickback if it is in the $\ket{1}$ state and will remain unchanged if it is in the $\ket{0}$ state, which means that the gate needed for that operation is a $CNOT$ using the i-th qubit of the second register as the control qubit and $\ket{-}$ as the target qubit. Therefore if the values of $g(s_{in})$ fulfil $g(s_{in})=s \cdot s_{in}$ then $G$ can be implemented only using $CNOT$s, fulfilling one of the requirements of a \textit{T-lineal circuit with reduced circuit complexity (RCC)}.
  	
  	Another way to reach the same conclusion is to notice that	the oracle $G$ in equation \eqref{G} turns into the original oracle of the non-recursive BV problem in equation \eqref{oracle1} if $g(s_{in})$ fulfils $ g(s_{in})=s \cdot s_{in}$. The oracle of the non-recursive BV problem is implemented with just $CNOT$s, so the oracle $G$ can also be implemented using only $CNOT$s. This property is also valid for any $k$ because $G$ only acts on one register at a time.

  	Next we want to prove that the values of $s_{x_1}$ from lemma \ref{def1} lead to implementing $U_s$ with Toffoli gates. The Toffoli gate performs the mapping $TOFF\ket{a,b,c}$=$\ket{a,b,c \oplus (a \wedge b)}$. Then if a Toffoli is applied, where the control qubits are one qubit from the first register and another from the second register, a phase kickback will occur when both qubits are in the $\ket{1}$ state and the target qubit is in the $\ket{-}$ state:
  	\begin{equation*}
  	TOFF\ket{1,1,-}=\ket{1,1,\left(\frac{\ket{0}-\ket{1}}{\sqrt{2}}\right)  \oplus (1 \wedge 1)}=-\ket{1,1,-}.
  	\end{equation*}
  	For the values of each $s_{x_1}$, the bitwise AND operation is performed using $x_1$ and $s$ as inputs, $s_{x_{1}}=x_1 \wedge s$  $\forall x_1$. This is equivalent to taking the corresponding $x_1$ for each $s_{x_1}$ and comparing it with $s$.  $s_{x_1}$ will have a 1 in every position where $x_1$ and $s$ both have a 1. Then:

  	\begin{enumerate}
  		\item   For the values of $s_{x_1}$ obtained from lemma \ref{def1}, every 1 that each $s_{x_1}$ has in the i-th position is also present in the i-th position of the corresponding $x_1$.
  		\item If $(-1)^{s_{x_1} \cdot x_2}=-1$ then this means that $x_2$ has a 1 in the same position $s_{x_1}$ has a 1. 
  		\item Then combining the two previous items, we have that $x_2$ has a 1 in the same position $x_1$ has a 1.
  		\item This means the operation that has to be implemented is a phase kickback if $x_1$ and $x_2$ have both a 1 in the i-th position, which is the Toffoli gate.
  	\end{enumerate}
  	Therefore if $s_{x_1}$ has a 1 in the i-th position, a Toffoli gate whose control qubits are those in the i-th position in each register will be applied in order to implement $(-1)^{s_{x_1} \cdot x_2}$ for each value of $s_{x_1}$. The same Toffoli applied to the i-th qubit will be used for every $s_{x_1}$ that has a 1 in the i-th position. Then the oracle $U_s$ can be implemented using only Toffoli gates. Finally since every 1 that each $s_{x_1}$ has in the i-th position is also present in $s$ then for every 1 in $s$ a Toffoli is needed. There are $n$ bits in $s$ so at most $n$ Toffoli gates can be applied. This concludes the proof of lemma \ref{def1}.
  \end{proof}
    
The values of $s_{x_1}$ obtained from lemma \ref{def1} are valid for a given $s$ because they are guaranteed to return the correct value when $g(s_{x_1})$ is applied later, since in both the operations used to obtain $s_{x_1}$ (the operation $s\wedge x_1)$ and $g(s_{in})$ (the operation $s \cdot s_{in})$ the number of 1s and the positions of these 1s in $s$, $x_1$ and $s_{in}$ are compared, the only difference is that the first one outputs another string and the latter outputs a 1 or a 0.

Notice that $G$ can be implemented using only $CNOT$s if half the values obey $g(s_{in})=0$ and the other half of the values obey $g(s_{in})=1$. This is the case for \textit{ T-lineal circuits with reduced circuit complexity (RCC)}. If $G$ has more values where $g(s_{in})$ equals 0 than those that equal 1, or viceversa, $CNOT$s are not enough because these gates would apply phase kickbacks not only to the correct states but also to some of the states that do not need phase kickbacks, since for $2^n$ states a single bit is 0 for half of the states and 1 for the other half.
\begin{figure}[]
	\includegraphics[width=1\columnwidth]{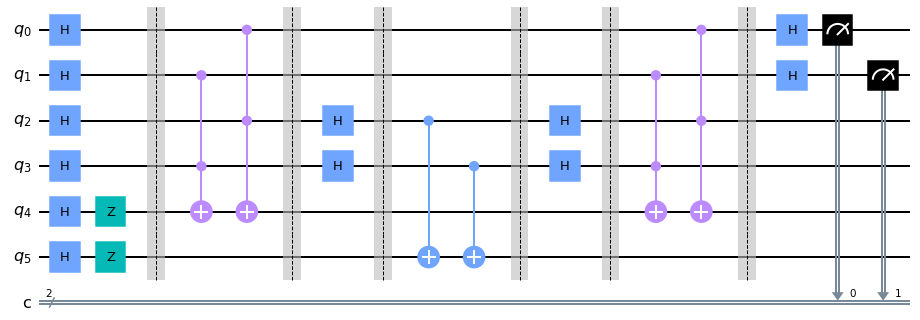}
	\centering
	\caption{Recursive Bernstein-Vazirani circuit for two recursions, $s={11}$, $s_{00}=00$, $s_{01}=01$, $s_{10}=10$ and $s_{11}=11$, $g(00)=g(11)=0$ and $g(01)=g(10)=1$. This circuit is obtained from lemma \ref{def1}.}
	\label{bv11toffoli}
\end{figure}

As an example of this type of circuits, we solve a 2 qubits circuit, where $s=11$. Since $g(s_{in})$ fulfils $g(s_{in})=s \cdot s_{in}$, we have $g(00)=g(11)=0$ and $g(01)=g(10)=1$. For $s_{x_1}$, we have $s_{x_1}=x_1 \wedge s$ $\forall x_1$ so : $s_{00}=00$, $s_{01}=01$, $s_{10}=10$, $s_{11}=11$. Since these values were obtained from lemma \ref{def1} then a \textit{T-lineal circuit with reduced circuit complexity (RCC)} can be constructed, so this circuit can be implemented using just $n=2$ Toffoli gates for the oracle $U_s$ and $CNOT$s for the oracle $G$. The circuit that completes this process represented in qiskit is shown in figure \ref{bv11toffoli}. The qubits are denoted by $q_0$, $q_1$,..., $q_{n-1}$, where $q_0$ refers to the first one, $q_1$ to the second one and $q_{n-1}$ to the $n$-th one. In this example, the first register has two qubits, $q_0$ and $q_1$, and the second register has qubits $q_2$ and $q_3$. 

We start by applying $U_s$ to the superposition of states of $x_1$ and $x_2$ in equation \eqref{superposition}. Since $s_{01}=01$, the second qubits of each register, $q_1$ and $q_3$, are used as the control qubits for a Toffoli gate whose target is the first extra qubit in the $\ket{-}$ state. This first Toffoli is denoted as $TOFF_{q_1,q_3,q_4}$ where $q_1$ and $q_3$ are control qubits and $q_4$ is the target qubit. Applying this Toffoli gives us the state (see figure \ref{bv11toffoli}): 
\begin{flalign*}
&TOFF_{q_1,q_3,q_4}\Big[\frac{1}{4}\Big[\ket{00}_1\otimes(\ket{00}_2+\ket{01}_2+\ket{10}_2+\ket{11}_2)&&\\\nonumber
&+\ket{01}_1\otimes(\ket{00}_2+\ket{01}_2+\ket{10}_2+\ket{11}_2)&&\\\nonumber
&+\ket{10}_1\otimes(\ket{00}_2+\ket{01}_2+\ket{10}_2+\ket{11}_2)&&\\\nonumber
&+\ket{11}_1\otimes(\ket{00}_2+\ket{01}_2+\ket{10}_2+\ket{11}_2)\Big]\otimes \ket{-}\Big].&&\\\nonumber
&=\frac{1}{4}\Big[\ket{00}_1\otimes(\ket{00}_2+\ket{01}_2+\ket{10}_2+\ket{11}_2)&&\\\nonumber
&+\ket{01}_1\otimes(\ket{00}_2-\ket{01}_2+\ket{10}_2-\ket{11}_2)&&\\\nonumber
&+\ket{10}_1\otimes(\ket{00}_2+\ket{01}_2+\ket{10}_2+\ket{11}_2)&&\\\nonumber
&+\ket{11}_1\otimes(\ket{00}_2-\ket{01}_2+\ket{10}_2-\ket{11}_2)\Big]\otimes \ket{-}.
\end{flalign*}

The next $s_{x_1}$ is $s_{10}=10$, so the first qubit of each register is used as the control qubit for a Toffoli gate. This means the control qubits are $q_0$ and $q_2$. The second Toffoli transforms the state into:

\begin{flalign*}
&TOFF_{q_0,q_2,q_4}\Big[\frac{1}{4}\Big[\ket{00}_1\otimes(\ket{00}_2+\ket{01}_2+\ket{10}_2+\ket{11}_2)&&\\\nonumber
&+\ket{01}_1\otimes(\ket{00}_2-\ket{01}_2+\ket{10}_2-\ket{11}_2)&&\\\nonumber
&+\ket{10}_1\otimes(\ket{00}_2+\ket{01}_2+\ket{10}_2+\ket{11}_2)&&\\\nonumber
&+\ket{11}_1\otimes(\ket{00}_2-\ket{01}_2+\ket{10}_2-\ket{11}_2)\Big]\otimes \ket{-}\Big]=&&\\\nonumber
&=\frac{1}{4}\Big[\ket{00}_1\otimes(\ket{00}_2+\ket{01}_2+\ket{10}_2+\ket{11}_2)&&\\\nonumber &+\ket{01}_1\otimes(\ket{00}_2-\ket{01}_2+\ket{10}_2-\ket{11}_2)&&\\\nonumber
&+\ket{10}_1\otimes(\ket{00}_2+\ket{01}_2-\ket{10}_2- \ket{11}_2)&&\\\nonumber
&+\ket{11}_1\otimes(\ket{00}_2-\ket{01}_2-\ket{10}_2+\ket{11}_2)\Big]\otimes \ket{-}.
\end{flalign*}

The combination of both Toffolis also implements $s_{11}=11$, because $s_{11}$ has two 1s. Applying two $H$ gates to the second register transforms the state into:

\begin{flalign*}
&I\otimes H^2\Big[\frac{1}{4}[\ket{00}_1\otimes(\ket{00}_2+\ket{01}_2+\ket{10}_2+\ket{11}_2)&&\\\nonumber &+\ket{01}_1\otimes(\ket{00}_2-\ket{01}_2+\ket{10}_2-\ket{11}_2)&&\\\nonumber
&+\ket{10}_1\otimes(\ket{00}_2+\ket{01}_2-\ket{10}_2- \ket{11}_2)&&\\\nonumber
&+\ket{11}_1\otimes(\ket{00}_2-\ket{01}_2-\ket{10}_2+\ket{11}_2)]\otimes \ket{-}\Big]=&&\\\nonumber
&=\frac{1}{2}\Big[\ket{00}_1\otimes\ket{00}_2+\ket{01}_1\otimes\ket{01}_2
&&\\\nonumber &+\ket{10}_1\otimes\ket{10}_2+\ket{11}_1\otimes\ket{11}_2\Big]\otimes \ket{-}.
\end{flalign*}

Then, we apply $G$ using two CNOTs since $s=11$, which gives us the state (see figure \ref{bv11toffoli}):

\begin{flalign*}
&I \otimes G\Big[\frac{1}{2}[\ket{00}_1\otimes\ket{00}_2+\ket{01}_1\otimes\ket{01}_2
&&\\\nonumber
&+\ket{10}_1\otimes\ket{10}_2+\ket{11}_1\otimes\ket{11}_2]\otimes \ket{-} \otimes \ket{-}\Big]=&&\\\nonumber
&=\frac{1}{2}\Big[\ket{00}_1\otimes\ket{00}_2-\ket{01}_1 \otimes\ket{01}_2&&\\\nonumber
&-\ket{10}_1\otimes\ket{10}_2+\ket{11}_1\otimes\ket{11}_2\Big]\otimes  \ket{-} \otimes \ket{-}.
\end{flalign*}

Since $g(01)=g(10)=1$ a phase kickback is applied to the states $\ket{01}_2$ and $\ket{10}_2$. Now we just need to apply $H^2$ and $U_s$ to the second register again. After two $H$ gates:

\begin{flalign*}
&I \otimes H^2\Big[\frac{1}{2}\Big[\ket{00}_1\otimes\ket{00}_2-\ket{01}_1 \otimes\ket{01}_2 &&\\\nonumber
&-\ket{10}_1\otimes\ket{10}_2+\ket{11}_1\otimes\ket{11}_2\Big]\otimes  \ket{-} \otimes \ket{-}\Big]=&&\\\nonumber &=\frac{1}{4}\Big[\ket{00}_1\otimes(\ket{00}_2+\ket{01}_2+\ket{10}_2+\ket{11}_2)&&\\\nonumber
&-\ket{01}_1\otimes(\ket{00}_2-\ket{01}_2+\ket{10}_2-\ket{11}_2)&&\\\nonumber
&-\ket{10}_1\otimes(\ket{00}_2+\ket{01}_2-\ket{10}_2- \ket{11}_2)&&\\\nonumber
&+\ket{11}_1\otimes(\ket{00}_2-\ket{01}_2-\ket{10}_2+\ket{11}_2)\Big]\otimes \ket{-}\otimes \ket{-}.
\end{flalign*}

After $U_s$ (the same two Toffolis as applied before):

\begin{flalign*}
&U_s\Big[\frac{1}{4}[\ket{00}_1\otimes(\ket{00}_2+\ket{01}_2+\ket{10}_2+\ket{11}_2)&&\\\nonumber
&-\ket{01}_1\otimes(\ket{00}_2-\ket{01}_2+\ket{10}_2-\ket{11}_2)&&\\\nonumber
&-\ket{10}_1\otimes(\ket{00}_2+\ket{01}_2-\ket{10}_2- \ket{11}_2)&&\\\nonumber
&+\ket{11}_1\otimes(\ket{00}_2-\ket{01}_2-\ket{10}_2+\ket{11}_2)]\otimes \ket{-}]\otimes \ket{-}\Big]=&&\\\nonumber
&=\frac{1}{4}\Big[(\ket{00}_1-\ket{01}_1-\ket{10}_1+\ket{11}_1)&&\\\nonumber
&\otimes(\ket{00}_2+\ket{01}_2+\ket{10}_2+\ket{11}_2)\Big]\otimes \ket{-} \otimes \ket{-}.
\end{flalign*}

As always, applying $H$ gates to the first register and measuring it will give us the state $\ket{11}_1$ which is the value of $s$.

Notice that if a given value of $s$ has a Hamming weight (the number of 1s contained in the bit string, denoted by $|x|$) lower than $n$, that value of $s$ must contain some 0s in its entries. Then since the result of the product $g(s_{in})=s\cdot s_{in}$ is independent of the values of the input in the positions where $s$ is 0, the values of each $s_{x_1}$ in the positions where $s$ is a 0 are irrelevant. These values can be a 0 or a 1 because it does not affect the result of the product and then the values of each $g(s_{x_1})$ are still correct. This means that for a circuit obtained from lemma \ref{def1}, if $|s|<n$ then extra Toffolis can be applied in the positions where $s$ has its 0s. The resulting circuit is still valid. This leads to lemma \ref{def2}:

\begin{tcolorbox}
	[colback=white]
	\begin{lem}
		Given a function $g:\{0,1\}^n\rightarrow \{0,1\}$ on $s_{in}$, $g(s_{in})$, that also obeys $g(s_{x_1})=s \cdot x_1$ where $s \in \{0,1\}^n$.
		For a $n$ bits string $s_{sup}$ $\in \{0,1\}^n$ that obeys $n\geq|s_{sup}|>|s|$ and $s_{sup} \wedge s=s$, where $|s|$ is the Hamming weight of the bit string, then a T-lineal circuit with reduced circuit complexity (RCC) can be constructed in which $g(s_{in})=s \cdot s_{in}$ and $s_{x_{1}}=x_1 \wedge s_{sup}$  $\forall x_1$ where  $x_1 \wedge  s_{sup}$ represents the bitwise AND operation and $x_1$ is a string, $x_1 \in \{0,1\}^n$.
		\label{def2}
	\end{lem}
\end{tcolorbox}
\begin{proof}
	The values of $g(s_{in})$ are the same as in lemma \ref{def1} so $G$ is implemented using $CNOT$ gates. Each $s_{x_1}$ is obtained from a bitwise AND operation between $x_1$ and another $n$ bits string like in lemma \ref{def1} so $U_s$ is implemented using Toffoli gates. Then we only need to prove that the values of each $s_{x_1}$ obtained from $s_{sup}$ are valid:

	\begin{enumerate}
		\item $s_{sup}$ obeys $n\geq|s_{sup}|>|s|$ so it contains a larger number of 1s than $s$.
		\item $s_{sup}$ also follows $s_{sup} \wedge s=s$, so it contains all the 1s already present in $s$.
		\item \label{item3} The result of the product  $g(s_{in})=s\cdot s_{in}$ is independent of the values of the input in the same positions where $s$ has a 0.
		\item All the 1s each $s_{x_1}$ contains are also present in $s_{sup}$. Then combining the previous items all the 1s each $s_{x_1}$ contains are also present in $s$ or they are in a position where $s$ is 0. Therefore, due to item 3, the values of each $g(s_{x_1})$ remain the same for the values of $s_{x_1}$ obtained from $s_{sup}$ than for those obtained from $s$ like in lemma \ref{def1}.
		
	\end{enumerate}
Therefore if $|s|< n$, Toffoli gates can be applied in the qubits of both registers using as control qubits those in the same positions where $s$ has its 0s. The resulting circuit is correct as long as Toffolis were applied to the positions where $s$ has a 1 like in lemma \ref{def1}. This concludes the proof of lemma \ref{def2}.
\end{proof}

\begin{figure}[]
	\includegraphics[width=1\columnwidth]{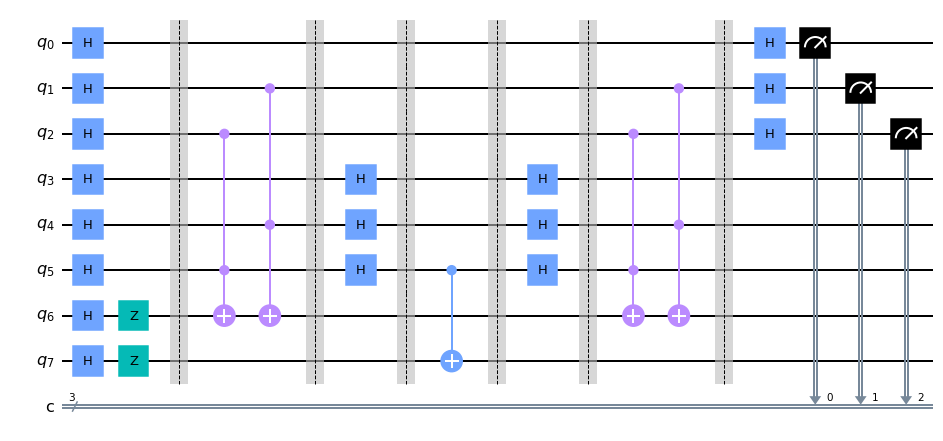}
	\centering
	\caption{Recursive Bernstein-Vazirani circuit for two recursions, $s={001}$, $s_{000}=000$, $s_{001}=001$, $s_{010}=010$, $s_{011}=011$, $s_{100}=000$, $s_{101}=001$, $s_{110}=010$, $s_{111}=011$ and $g(000)=g(010)=g(100)=g(110)=0$, $g(001)=g(011)=g(101)=g(111)=1$. This circuit is obtained from lemma \ref{def2}.}
	\label{bv011_2tof}
\end{figure}

As an example, if $s=001$, then the available values for $s_{sup}=\{011,101,111\}$ due to the condition $s_{sup}\wedge s=s$.
The values of $s_{x_{1}}$ for $s_{sup}=011$ according to lemma \ref{def2} would be $s_{x_{1}}=x_1 \wedge s_{sup}$ $\forall x_1$ so: $s_{000}=000$, $s_{001}=001$, $s_{010}=010$, $s_{011}=011$, $s_{100}=000$, $s_{101}=001$, $s_{110}=010$, $s_{111}=011$. Since $g(s_{in})$ fulfils $g(s_{in})=s \cdot s_{in}$ the values of $g(s_{in})$ would be $g(000)=g(010)=g(100)=g(110)=0$, $g(001)=g(011)=g(101)=g(111)=1$, so the values of $g(s_{in})$ in this circuit are the same as if they were in the $s=001$ case but with different values of $s_{x_{1}}$. For these values a \textit{T-lineal circuit with reduced circuit complexity (RCC)} can be constructed. The circuit that solves this example is represented in figure \ref{bv011_2tof}. Notice that $U_s$ is made using two Toffolis (even though $s$ only has a 1) but $G$ is constructed with just one $CNOT$.

The possible permutations of the control qubits of the second register of the Toffolis also result in \textit{T-lineal circuits with reduced circuit complexity (RCC)}. For a valid circuit obtained from lemma \ref{def1} with at least two Toffolis where the number of Toffoli gates equals the Hamming weight of the corresponding bit string $s$ of the circuit, $|s|$, permuting the controls applied to the second register of these gates also produces a \textit{T-lineal circuit with reduced circuit complexity (RCC)}. The permuted valid values of $s_{x_1}$ are obtained by permuting the bits of the values of each $s_{x_1}$ obtained from lemma \ref{def1}. The bits that can be permuted are those where $s$ is 1 (likewise the control qubits of the second register allowed to be permuted are those in the same positions as the 1s in $s$).  

\begin{figure}[]
 	\includegraphics[width=1\columnwidth]{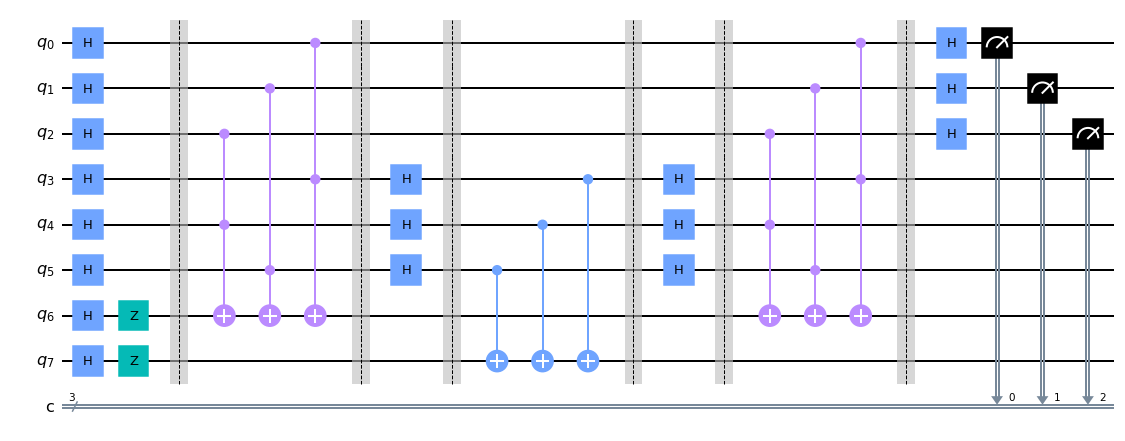}
 	\centering
 	\caption{Recursive Bernstein-Vazirani circuit for two recursions, $s=111$, $s_{000}=000$, $s_{001}=010$, $s_{010}=001$, $s_{011}=011$, $s_{100}=100$, $s_{101}=110$, $s_{110}=101$, $s_{111}=111$  and $g(000)=g(011)=g(101)=g(110)=0$, $g(001)=g(010)=g(100)=g(111)=1$. This circuit is obtained from lemma \ref{def1} after permuting the second and third control qubits of the second register.}
 	\label{bv_permutation}
 \end{figure}

For a $n$ bits string $s$ with Hamming weight $|s|$ the number of possible permutations of $s_{x_1}$ that lead to \textit{T-lineal circuit with reduced circuit complexity (RCC)} are $|s|!-1$ (all the possible permutations except the one already given by lemma \ref{def1}). As an example, if $s=111$, then there are $3!-1$ possible permutations available to the values of $s_{x_1}$ obtained from lemma \ref{def1}. Starting with the values obtained from lemma \ref{def1}: $s_{000}=000$, $s_{001}=001$, $s_{010}=010$, $s_{011}=011$, $s_{100}=100$, $s_{101}=101$, $s_{110}=110$, $s_{111}=111$. The first permutation could be switching the position of bits 2 and 3:  $s_{000}=000$, $s_{001}=010$, $s_{010}=001$, $s_{011}=011$, $s_{100}=100$, $s_{101}=110$, $s_{110}=101$, $s_{111}=111$. This means that the control qubits 2 and 3 of the second register of the Toffolis will be permuted, as can be seen in figure \ref{bv_permutation} which shows the circuit that solves this example. The rest of the permuted circuits can be obtained using this procedure.  This process works because the permuted values of $s_{x_1}$
 have the same values for $g(s_{x_1})$ before and after the permutations. Since the permuted bits of each $s_{x_1}$ are those in which $s$ has a 1, once $g(s_{in})=s \cdot s_{in}$ is applied all the permuted bits will be part of a modulo 2 sum. This sum commutes so the permuted bits are allowed to change positions and leave the result of each $g(s_{x_1})$ unchanged.
 
Using the same control qubit of the second register for all the Toffolis for a given circuit obtained from lemma \ref{def1} also results in a valid \textit{T-lineal circuit with reduced circuit complexity (RCC)}.
In this case the values of each $s_{x_1}$ would be the bit string where all the bits are 0 for those states where $g(s_{x_1})=s \cdot x_1=0$ and the bit string where all bits are 0 but 1 for the states where $g(s_{x_1})=s \cdot x_1=1$. The position of this 1 would be the same position of the control qubit of the second register that all the Toffolis use. The values of $s_{x_1}$ would be: $s_{x_1}=s_0$ if $s \cdot x_1=0$ and $s_{x_1}=s_1$ if $s \cdot x_1=1$, where $s_0$ is a $n$ bits string with $|s_0|=0$ and $s_1$ is a $n$ bits string with $|s_1|=1$ (all 0s except just a 1). This leads to lemma \ref{def3} :

\begin{tcolorbox}
	[colback=white]
	\begin{lem}
		Given a function $g:\{0,1\}^n\rightarrow \{0,1\}$ on $s_{in}$, $g(s_{in})$, that also obeys $g(s_{x_1})=s \cdot x_1$.
		For a $n$ bits string $s$,  $s \in \{0,1\}^n$, and two $n$ bits strings $s_0$ and $s_1$ which obey $|s_0|=0$ and $|s_1|=1$ where $|s|$ is the Hamming weight of $s$, then a T-lineal circuit with reduced circuit complexity (RCC) can be constructed in which $|s\wedge s_1|=1$; $g(s_{in})=s \cdot s_{in}$; $s_{x_{1}}=s_0 $ $\forall s_{x_1}$ if $s \cdot x_1=0$ and  $s_{x_1}=s_1$ $\forall s_{x_1}$ if $s \cdot x_1=1$ where $x_1$ is a string, $x_1 \in \{0,1\}^n$.
		\label{def3}
	\end{lem}
\end{tcolorbox}

\begin{proof}
	We only need to prove that the values for $s_{x_1}$ can be implemented using Toffoli gates with the same qubit from the second register as the control qubit.  
	\begin{enumerate}
		\item   For the non zero values of $s_{x_1}$ obtained from lemma \ref{def3} we have that $s_{x_1}=s_1$. Then every $s_{x_1}$ has a 1 in the same fixed i-th position. The condition $|s\wedge s_1|=1$ ensures that the 1 in $s_{x_1}$ is in the same position $s$ has a 1.
		\item If $(-1)^{s_{x_1} \cdot x_2}=(-1)^{s_{1} \cdot x_2}=-1$ then this means that the phase kickback will activate only when $x_2$ has a 1 in the same fixed position as $s_{x_1}$. This only happens when $g(s_{x_1})=s \cdot x_1=1$.
		\item Since $s_{x_1}=s_1$ only if $g(s_{x_1})=s \cdot x_1=1$, then $x_1$ has a 1 in the same position $s$ has a 1.
		\item Combining the previous items then the operation that has to be implemented is a phase kickback if $x_1$ has a 1 in the same position $s$ has a 1 and $x_2$ has a 1 in the same fixed position $s_{x_1}$ has a 1, which is a Toffoli gate.	
	\end{enumerate}
	Therefore if $s$ has a 1 in the i-th position, a Toffoli gate whose first control qubit is in the i-th position in the first register and whose second control qubit is always in the same position in the second register will be applied in order to implement $(-1)^{s_{x_1} \cdot x_2}$ for each value of $s_{x_1}$.
	For each 1 in $s$ a Toffoli gate is required so if $|s|=n$, $n$ of these Toffoli gates are used. When an even number of Toffolis activate since all act on the same qubit of the second register they cancel each other and when an odd number activates, $(-1)^{s_1 \cdot x_2}$ will be implemented. This concludes the proof of lemma \ref{def3}.
\end{proof}

\begin{figure}[]
	\includegraphics[width=1\columnwidth]{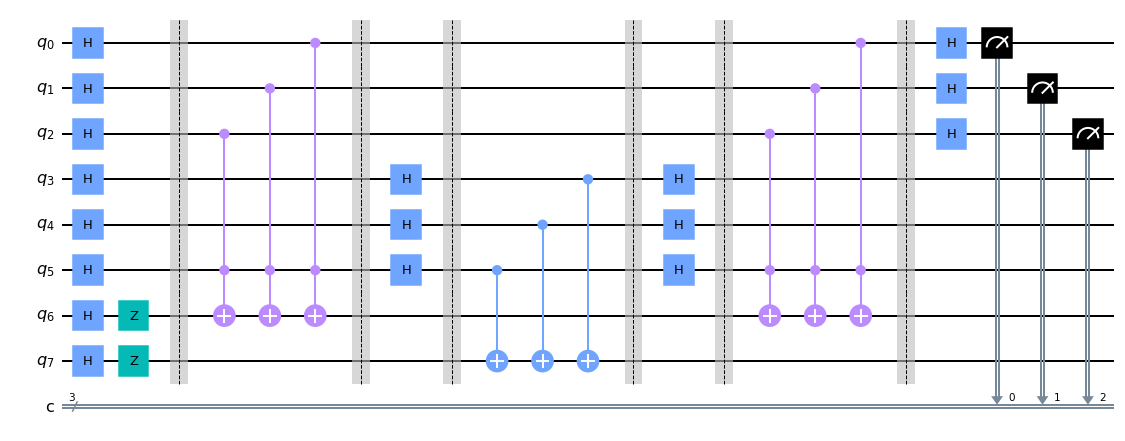}
	\centering
	\caption{Recursive Bernstein-Vazirani circuit for two recursions, $s=111$, $s_{000}=000$, $s_{001}=001$, $s_{010}=000$, $s_{011}=001$, $s_{100}=000$, $s_{101}=001$, $s_{110}=000$, $s_{111}=001$  and $g(000)=g(011)=g(101)=g(110)=0$, $g(001)=g(010)=g(100)=g(111)=1$. This circuit is obtained from lemma \ref{def3}.}
	\label{bv_control}
\end{figure}

As an example, if $s=111$, then $s_{000}=000$, $s_{001}=001$, $s_{010}=000$, $s_{011}=001$, $s_{100}=000$, $s_{101}=001$, $s_{110}=000$, $s_{111}=001$, so $s_1=001$. The control qubit of the second register for all Toffolis is the third qubit.  The resulting circuit is represented in figure \ref{bv_control}.

Toffoli gates where the control qubit of the first register activates when the qubit is in the $\ket{0}$ state instead of the $\ket{1}$ state also lead to valid circuits, although only for values of $s$ with an even number of 1s (assuming that $g(s_{in})$ fulfils $g(s_{in})=s\cdot s_{in}$). This is because for a $s$ with odd Hamming weight $s_{11...11}= s_0$ is impossible since $g(s_0)=0$ and $g(s_{11...11})=s\cdot 11...11=1$. When the first register is in the $\ket{11...11}$ state, applying a phase kickback to the second register is only possible with controlled gates where the controls work as usual (they activate when the control is a 1). If $|s|$ is even, $s_{11...11}=s_0$ is a valid value since $g(s_{11...11})=s\cdot 11...11=0$ for a $s$ with even Hamming weight, so when the first register is in the $\ket{11...11}$ state, there is no need for $U_s$ to apply a phase kickback to the second register to implement $(-1)^{s_{11...11}\cdot x_2}$. If the Toffolis used are then negated on the first register's qubits, \textit{ T-lineal circuits with reduced circuit complexity (RCC)} can be constructed where $s_{x_1}=s \wedge \overline{x_1}$ ($\overline{x_1}$ is the conjugate of $x_1$) and $g(s_{in})$ fulfils $g(s_{in})=s\cdot s_{in}$ as usual. The position of the control qubits of the Toffolis for both registers would be the same as the position of each 1 in $s$ as in lemma \ref{def1}. This leads to lemma \ref{def4}:

\begin{tcolorbox}
	[colback=white]
	\begin{lem}
		Given a function $g:\{0,1\}^n\rightarrow \{0,1\}$ on $s_{in}$, $g(s_{in})$, that also obeys $g(s_{x_1})=s \cdot x_1$.
		For a $n$ bits string $s$,  $s \in \{0,1\}^n$, for which $|s|$ is even where $|s|$ is the Hamming weight of $s$, then a T-lineal circuit with reduced circuit complexity (RCC) can be constructed in which $g(s_{in})=s \cdot s_{in}$, $s_{x_{1}}=s \wedge \overline{x_1} $ $\forall s_{x_1}$ where $s \wedge \overline{x_1}$  represents the bitwise AND operation between  $\overline{x_1}$ and $s$, $x_1$ is a string, $x_1 \in \{0,1\}^n$ and $\overline{x_1}$ is the conjugate of $x_1$.
		\label{def4}
	\end{lem}
\end{tcolorbox}

\begin{proof}
	We only need to prove that the values for $s_{x_1}$ can be implemented using Toffoli gates with the negated control qubits in the first register. 
	\begin{enumerate}
		\item   For the values of $s_{x_1}$ obtained from lemma \ref{def4}, every 1 that each $s_{x_1}$ has in the i-th position corresponds to a 0 in the i-th position of the respective $x_1$.
		\item If $(-1)^{s_{x_1} \cdot x_2}=-1$ then this means that $x_2$ has a 1 in the same position $s_{x_1}$ has a 1. 
		\item Then combining the two previous items, we have that $x_2$ has a 1 in the same position $x_1$ has a 0.
		\item This means the operation that has to be implemented is a phase kickback if $x_1$ has a 0 and $x_2$ has a 1 in the i-th position, which is the Toffoli gate with negated control qubits in the first register.
	\end{enumerate}
	Therefore if $s_{x_1}$ has a 1 in the i-th position, a Toffoli gate whose control qubits are those in the i-th position in each register will be applied in order to implement $(-1)^{s_{x_1} \cdot x_2}$ for each value of $s_{x_1}$. The control qubits of the first register are negated. The same Toffoli applied to the i-th qubit will be used for every $s_{x_1}$ that has a 1 in the i-th position. This concludes the proof of lemma \ref{def4}.
\end{proof}

\begin{figure}[]
	\includegraphics[width=1\columnwidth, height=0.4\linewidth]{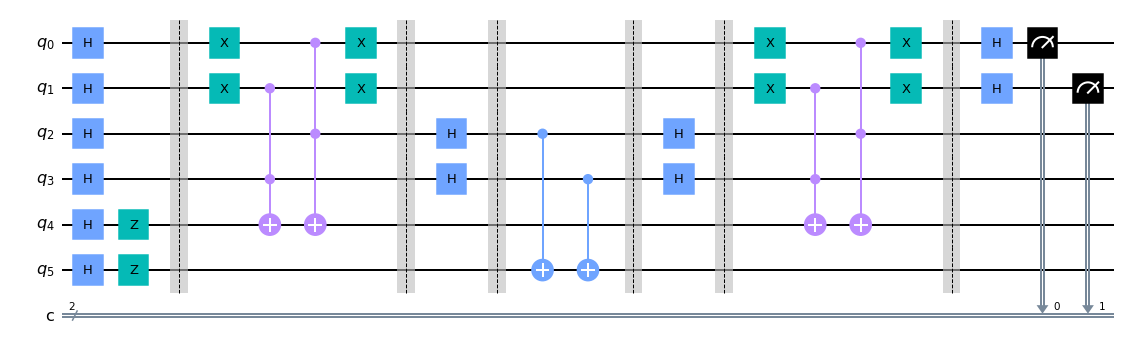}
	\centering
	\caption{Recursive Bernstein-Vazirani circuit for two recursions,  $s=11$, then $s_{00}=11$, $s_{01}=10$, $s_{10}=01$, $s_{11}=00$ and $g(00)=g(11)=0$, $g(01)=g(10)=1$. This circuit is obtained from lemma \ref{def4}.}
	\label{bv_neg}
\end{figure}

As an example, if $s=11$, then $s_{00}=11$, $s_{01}=10$, $s_{10}=01$, $s_{11}=00$ and the values of $g(s_{in})$ would be $g(00)=g(11)=0$, $g(01)=g(10)=1$. The circuit that solves the BV problem for these values is represented in figure \ref{bv_neg}. As it was explained, the control qubits of the first register of the Toffolis are surrounded by $X$ gates, so they activate when their state is a 0 instead of a 1.

What was discussed about the permutations of control qubits of the Toffolis of the second register works in the same manner for these circuits where negated Toffolis are used. All possible circuits are obtained permuting the control qubits in the same positions where $s$ has 1s. This is because as before the permuted values of $s_{x_1}$ have the same values for $g(s_{x_1})$ before and after the permutations.

The circuits obtained from the values of $s_{x_1}$ and $g(s_{in})$ given by lemma \ref{def1} along with their corresponding permutations, the circuits obtained from lemma \ref{def2}, \ref{def3} and those obtained from lemma \ref{def4} (along with their valid permutations) constitute the complete set of \textit{ T-lineal circuits with reduced circuit complexity (RCC)} for $k=2$.

Regarding the number of Toffoli gates required to solve the BV problem for circuits obtained from lemma \ref{def1}, in the worst case scenario $n$ gates are needed, since for a $s$ with $n$ bits where $|s|=n$ all qubits in both registers will require a Toffoli gate to perform the necessary phase kickbacks. This is because if $|s|=n$ then $s_{x_{1}}=x_1 \wedge s=x_1$ so each qubit in the second register will require a phase kickback. The best case scenario on the other hand requires just one Toffoli gate, which happens when $|s|=1$. If $s$ contains a 1 just once each $s_{x_{1}}$ will contain its 1 in the same position, which is the same position as the control qubits of both registers needed to implement the Toffoli gate.
The worst case scenario for the remaining possible circuits not obtained from lemma \ref{def1} still requires $n$ Toffoli gates to be implemented.

The main feature of these circuits is the fact that the oracle $U_s$ was simplified from multi-controlled $CNOT$ gates to just $n$ Toffoli gates at most and the oracle $G$ was also simplified so it could be implemented using $CNOT$ gates instead of multi-controlled $CNOT$ gates. The importance of this idea is related to the number of $T$ gates required to implement each operation.

\begin{figure}[]
	\includegraphics[width=0.9\columnwidth]{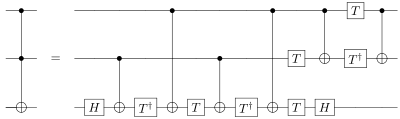}
	\centering
	\caption{A Toffoli gate requires 7 $T$ gates to be implemented.}
	\label{tofolli}
\end{figure}

For \textit{ T-lineal circuits with reduced circuit complexity (RCC)}, like in figure \ref{bv11toffoli}, the number of $T$ gates grows linearly with the number of qubits of each register (the same number as bits needed to represent $s$), $n$, required to solve the problem. This is because as it has already been discussed, in the worst case scenario $n$ Toffolis  are needed to implement $U_s$. Since a Toffoli can be decomposed in 7 $T$ gates as seen in figure \ref{tofolli} ($T^\dagger$ gates are counted as $T$ gates) and considering that $U_s$ has to be applied twice to solve the recursive BV problem for $k=2$, the number of $T$ gates needed is $14 \cdot n $, which means this number grows linearly with the number of qubits of each register $n$, so the number of $T$ gates grows as $O(n)$.

\begin{tcolorbox}
	[colback=white]
	\begin{cor}
		The number of $T$ gates needed to implement \textit{T-lineal circuits with reduced circuit complexity (RCC)} grows linearly with the number of qubits of each register needed to solve the problem: $O(n)$.
		\label{cor3}
	\end{cor}
\end{tcolorbox}

The importance of corolary \ref{cor3} will be understood once it is related to Quantum Homomorphic Encryption. 

On the other hand, for the case where $k=2$, if $U_s$ has to be implemented using just multi-controlled $CNOT$s, then the number of $T$ gates grows differently. Since $g(s_{x_1})=s \cdot x_1$, half of the values of $g(s_{x_1})$ will be 1, so half of all $s_{x_1}$s will have $|s|\neq 0$ (they will not be equal to $000...00$). This means that $U_s$ will require at least a multi-controlled $CNOT$ for half of the $s_{x_1}$s, because unlike Toffoli gates, a multi-controlled $CNOT$ is restricted to just one value of $x_1$ each time it is applied. Then, the number of multi-controlled $CNOT$s will be $2^{n-1}$ at minimum, since there are $2^n$ $s_{x_1}$s and we need to take half of that number. 

Considering that $U_s$ is applied twice, the number of multi-controlled $CNOT$s would be $2^n$. Even if for simplicity we assume that each multi-controlled $CNOT$ could be implemented using 7 $T$ gates like the Toffoli gates, the number of $T$ gates grows exponentially with the number of qubits of each register $n$, so the number of $T$ gates grows as $\Omega(2^n)$. This fact will also have consequences in the context of Quantum Homomorphic Encryption (this implementation of $U_s$ would not be suitable to be implemented using the quantum homomorphic encryption schemes that will be discussed). 

Assuming that the multi-controlled $CNOT$s have the same number of $T$ gates as the Toffoli gates is an extreme simplification, but since the number of $T$ gates is exponential even in this case, it is enough for our purpose: checking if the number of $T$ gates in the circuit is exponential or polynomial.

The states where $U_s$ has to be implemented with multi-controlled $CNOT$s (Toffolis are not enough) are those where the $s_{x_1}$s have large number of ones. If we define $s_{max}$ as the $n$ bits string where $|s_{max}|=n$, it is possible to construct a BV problem where half of the $s_{x_1}$s are equal to $s_{max}$ since $g(s_{max})=s \cdot s_{max}=0$ $\forall s$ if $|s|$ is even and $g(s_{max})=s \cdot s_{max}=1$ $\forall s$ if $|s|$ is odd (assuming $g(s_{in})$ fulfils $g(s_{in})=s \cdot s_{in}$).

As an example if $s=111$, then $s_{000}=000$, $s_{001}=111$, $s_{010}=111$, $s_{011}=011$, $s_{100}=111$, $s_{101}=101$, $s_{110}=110$, $s_{111}=111$ and $g(000)=g(011)=g(101)=g(110)=0$, $g(001)=g(010)=g(100)=g(111)=1$. Since $|s|$ is odd, $g(s_{max})=s \cdot s_{max}=s \cdot 111=1$ so $s_{max}$ is a valid value for all $s_{x_1}$s where $g(s_{x_1})=s \cdot x_1=1$. This example is basically what would be obtained from lemma \ref{def1} for $s=111$ but substituting half of the values of $s_{x_1}$ for $s_{max}$. However, it is impossible to implement this circuit just using Toffolis, because $s_{001}=s_{010}=s_{100}=111$, so when $U_s$ is applied, states where $g(s_{x_1})=s \cdot x_1=1$ like $\ket{001}_1$ in the first register would have to apply 3 phase kickbacks to the qubits of the second register. This would be impossible since $\ket{001}_1$ only contains a 1, so the first two qubits would not activate if a Toffoli was applied to them. Toffolis with negated controls could be applied but this would also change the remaining states, so this particular case of the problem can not be solved with Toffolis and needs multi-controlled CNOTs. In this particular example, applying 3 Toffolis would implement $U_s$ for all the $s_{x_1}$s where $g(s_{x_1})=0$ and would also implement $s_{111}=111$ (this is done to minimize the number of multi-controlled CNOTs as much as possible). However for the remaining values ($s_{001}$, $s_{010}$, $s_{100}$) two multi-controlled CNOTs would be needed for each $s_{x_1}$ so $U_s$ implements $s_{001}= s_{010}= s_{100}=s_{max}$. 

In the general case assuming that $n$ Toffolis are used so the maximum amount of phase kickbacks are applied, minimizing the number of multi-controlled CNOTs gates used in these circuits, $2^{n-1}$ states from the first register need to apply phase kickbacks to the second register because half of the $2^n$ $s_{x_1}$s equal $s_{max}$. Then all of these states except the $\ket{111...11}$ would need at least a multi-controlled $CNOT$ (most would need more) so at least $2^{n-1}-1$ multi-controlled $CNOT$s would be needed to implement $U_s$. Then, as it has been discussed, the best possible number of $T$ gates for these types of circuits grows as $\Omega(2^n)$. If no Toffolis were used, then still $2^{n-1}$ states would need phase kickbacks and each one would need $n$ phase kickbacks so $U_s$ implements $s_{max}$ for half of the $s_{x_1}$s, making the number of multi-controlled $CNOT$ gates in the circuit $2n*2^n$ (worse than the former $n$ Toffolis case).

We conclude this section with a possible application of \textit{T-lineal circuits with reduced circuit complexity (RCC)}. These circuits could be used as a test to asses that a quantum computer is functioning correctly. These circuits can not be efficiently simulated using classical computers since they contain $T$ gates but they require less computational power than circuits with exponential number of $T$ gates. We are currently in the so called ``noisy intermediate-scale quantum'' era (NISQ) in which the best quantum computers available have at most a few hundred qubits but still have not reached fault-tolerance and are still not large enough to make use of the full computational power quantum mechanics offer. Since the number of $T$ gates grows as $O(n)$ for \textit{T-lineal circuits with reduced circuit complexity (RCC)}, these circuits could be implemented in quantum computers with large number of qubits to certify their correct functioning by making sure the correct $s$ is returned and it would be less computationally difficult than other algorithms where the number of $T$ gates is higher than lineal. 

\section{Application to quantum homomorphic encryption}
\label{sec:application}

Our \textit{ T-lineal circuits with reduced circuit complexity (RCC)} would be perfect to be implemented using Liang's schemes, since as we have seen these circuits have a number of $T$ gates that grows linearly with the number of qubits of each register of the circuit $n$, so the number of $T$ gates grows as $O(n)$.

On the next sections we will briefly explain how the QHE scheme is constructed.

\subsection{Preliminaries}
In the circuit model of quantum computation, quantum gates are the basis of any algorithm that could be implemented \cite{articulo MA}. We will use the usual Clifford gates, which are $\{X,Z,H,S,CNOT\}$. $X=\begin{pmatrix}
0& 1 \\
1 & 0
\end{pmatrix}$ and $Z=\begin{pmatrix}
1& 0 \\
0 & -1
\end{pmatrix} $ are the usual Pauli gates. The Hadamard gate is $H=\frac{X+Z}{\sqrt{2}}=\frac{1}{\sqrt{2}}\begin{pmatrix}
1& 1 \\
1 & -1
\end{pmatrix}
$
. $S$ is just $\sqrt{Z}$ and $	CNOT=\begin{pmatrix}
1& 0&0&0 \\
0& 1&0&0 \\
0& 0&0&1 \\
0& 0&1&0 \\
\end{pmatrix}
$.
To perform universal quantum computation the $T= \begin{pmatrix}
1& 0 \\
0 & e^{i\frac{\pi}{4}}
\end{pmatrix}$ and $T^{\dagger}= \begin{pmatrix}
1& 0 \\
0 & e^{-i\frac{\pi}{4}}
\end{pmatrix}$ need to be added to the Clifford gates. The set of gates then becomes $\mathcal{S}= \{X,Z,H,S,CNOT,T,T^{\dagger}\}$. The main difficulty of the homomorphic evaluation of a quantum circuit consists on the evaluation of the $T$ and $T^{\dagger}$ gates because it causes a $S$-error: $TX^aZ^b\ket{\phi}=(S^{\dagger})^aX^aZ^{a\oplus b}T\ket{\phi}
$, so somehow this error has to be corrected. Liang's schemes correct it by making use of gate teleportation, which is a generalization of quantum teleportation.

\subsection{Gate teleportation}
An EPR state is an entangled quantum state $\ket{\Phi_{00}}=\frac{1}{\sqrt{2}} (\ket{00}+\ket{11})$.
From this entangled state, we can express the usual four Bell states in a compact expression:
\begin{equation}
\ket{\Phi_{ab}}=(Z^bX^a \otimes I)\ket{\Phi_{00}}, \forall a, b \in {0,1}.
\end{equation}
The state $\ket{\Phi_{00}}$ can be constructed by using a $H$ gate followed by a $CNOT$. Quantum teleportation is a procedure that transports quantum states between a sender and a receiver using a quantum communication channel. Alice wants to send Bob a state $\ket{\psi}$, so they share an entangled pair $\ket{\Phi_{00}}$ so each one has a qubit from the EPR pair. Alice makes a measurement in the Bell basis using her two qubits, $\ket{\psi}$ and one qubit of $\ket{\Phi_{00}}$. Due to entanglement, if Bob applies the correct quantum gates (a combination of $X$ and $Z$), he can reconstruct $\ket{\psi}$ \cite{teleportation}.

For any single gate $U$, the ``$U$-rotated Bell basis'' is defined as: $\Phi(U)=\{\ket{\Phi(U)_{ab}}, a,b \in \{0,1\} \}$, where $\ket{\Phi(U)_{ab}}=(U^{\dagger} \otimes I)\ket{\Phi_{ab}}=(U^{\dagger}Z^bX^a \otimes I)\ket{\Phi_{00}}$. 

For a single qubit we have the following expression from \cite{U rotated}, which is just an expression for quantum teleportation:
\begin{equation}
\ket{\alpha} \otimes \ket{\Phi_{00}} = \sum_{a,b\in \{0,1\}} \ket{\Phi_{ab}} \otimes X^aZ^b \ket{\alpha}
\end{equation}
It can be extended pretty easily for the ``$U$-rotated Bell basis'':
\begin{equation}
\ket{\alpha} \otimes \ket{\Phi_{00}} = \sum_{a,b\in \{0,1\}} \ket{\Phi(U)_{ab}} \otimes X^aZ^bU \ket{\alpha}
\label{gate teleport}
\end{equation}
where $U$ is any single qubit gate.

\begin{figure}[]
	\includegraphics[width=0.4\textwidth]{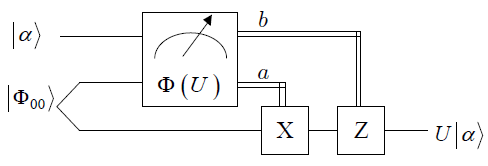}
	\centering
	\caption{Quantum circuit for ``gate teleportation''. The box represents the quantum measurement Alice performed on $\ket{\alpha}$ and one qubit of the pair $\ket{\Phi_{00}}$. The measurement basis is the $U$- rotated Bell basis $\Phi(U)$. Depending on the results of the measurement, $a$ and $b$, Bob applies $X^{a}$ and $Z^{b}$ in order to obtain $U \ket{\alpha}$.
	}
	\label{gatetelepory}
\end{figure}

Equation $\eqref{gate teleport}$ then describes ``gate teleportation''. Alice and Bob first share an entangled EPR pair $\ket{\Phi_{00}}$, then Alice prepares a state $\ket{\alpha}$ and performs a ``$U$-rotated Bell measurement''. The $U$-rotated Bell basis is selected as the measurement basis in the quantum measurement on the two qubits she has, $\ket{\alpha}$ and one of the pair $\ket{\Phi_{00}}$. She obtains the results $a$ and $b$ from the measurement, and Bob's qubit transforms into the state $X^aZ^bU \ket{\alpha}$, so once Alice tells Bob the results of her measurement, Bob can apply the Pauli $X$ and $Z$ operators as necessary to obtain $U \ket{\alpha}$. ``Gate teleportation'' is represented in figure \ref{gatetelepory}.

Gate teleportation is an extension of quantum teleportation since if $U$ were
the identity then $\Phi(U)$ would be the standard Bell basis, and gate teleportation
would be the standard quantum teleportation protocol.

This gate teleportation idea is the basis of Liang's QHE schemes, since it allows the correction of the error that results from homomorphically evaluating a $T$ gate.

\subsection{Encryption and evaluation}
The first step of Liang's schemes is applying Quantum One Time Pad (QOTP) to the data that will be sent to the server. This means applying a combination of $X^a$ and $Z^b$ to each qubit, where $a$ and $b$ are random bits selected from $\{0,1\}$ which constitute the secret key of the client $sk$. If the plaintext data has $n$ qubits the secret key $sk$ has $2n$ bits $sk=(a_0,b_0),a_0,b_0\in\{0,1\}^n$. The $w$-th plaintext qubit is encrypted using the secret bits $(a_0(w); b_0(w))$ so: $\ket{\alpha} \rightarrow X^{a_0}Z^{b_0} \ket{\alpha}=\ket{\rho_0}$. QOTP was proposed by Boykin and Roychowdhury in \cite{one time pad}, where they proved that as long as $a$ and $b$ are randomly selected from $\{0,1\}$ and used only once, QOTP has perfect security.

\begin{figure}[]
	\includegraphics[width=0.5\textwidth]{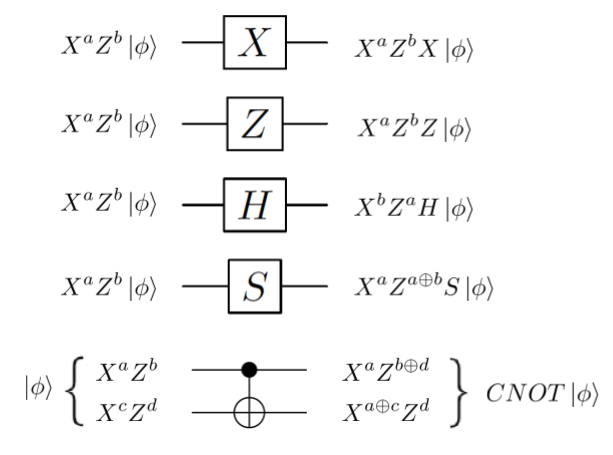}
	\centering
	\caption{Key updating rules for homomorphic evaluation of Clifford gates.
	}
	\label{reglas}
\end{figure}

Once the data has been encrypted it will be sent to the server. There, the server will perform quantum gates from the set $\mathcal{S}= \{X,Z,H,S,CNOT,T,T^{\dagger}\}$ to complete its designated quantum circuit. If the gates are not $T$ or $T^{\dagger}$ (just Clifford), the homomorphic evaluation con be performed easily. When the server performs one of these gates on a qubit, the key is updated according to the algorithm shown in figure \ref{reglas}. 
After the $j$th ($1\leq j \leq N-1$) gate is performed, a new key (denoted as
$(a_j, b_j), a_j , b_j \in\{0,1\}^n$) can be obtained through key updating; this is the intermediate key. As an example applying an $H$ gate would transform the key of the qubit as $(a_1,b_1)=(b_0,a_0)$ assuming it was the first gate in the circuit.

\begin{figure}[]
	\includegraphics[width=0.52\textwidth]{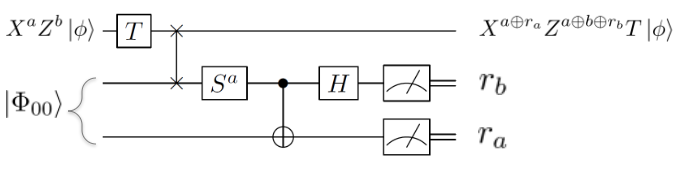}
	\caption{Homomorphic evaluation of $T$ gate. $\ket{\Phi_{00}}$ represents an EPR pair. 
	}
	\label{T gate}
\end{figure}

If the gate is a $T/T^{\dagger}$, its homomorphic evaluation is done using the $S^a$-rotated Bell measurement in order to correct the $S$-error that we have already mentioned. At the start of the evaluation of the whole circuit, an EPR source generates $M$ Bell states (as many as $T/T^{\dagger}$ gates are in the circuit) $\{\ket{\Phi_{00}}_{c_i,s_i}, i=1,..,M\}$, where qubits $c_i, i = 1,....,M$ and qubits $s_i, i =
1,....,M$ are kept by the client and the server respectively. When the server has to evaluate a $T/T^{\dagger}$ gate, it first applies the gate to a qubit, then performs a SWAP gate between the encrypted qubit and one of the entangled qubits the server has in its position, $s_i$. Then the only step remaining is to apply a $S^a$-rotated Bell measurement on the qubits $s_i$ and $c_i$. From this measurement, $r_a$ and $r_b$ will be obtained, so the encryption key can be updated. From $TX^aZ^b\ket{\alpha}$, this whole process returns $X^{a\oplus r_a}Z^{a \oplus b \oplus r_b}T\ket{\alpha}$. If the gate evaluated is a $T^{\dagger}$ gate the key will be updated from $T^{\dagger}X^aZ^b\ket{\alpha}$ to $X^{a\oplus r_a}Z^{ b \oplus r_b}T^{\dagger}\ket{\alpha}$. It is graphically represented in figure \ref{T gate}. This process will be performed $M$ times, as many as $T/T^{\dagger}$ gates are in the circuit.

Due to the principle of deferred measurement, the $M$ measurements can be postponed until the server has finished all the quantum operations. Then, the server sends all the encrypted qubits, the key-updating functions based on the rules of figure \ref{reglas} and all the ancillary $s_i$ qubits to the client. Then the client will update his keys accordingly and perform a measurement, alternating between updating and measuring, since to perform the correct $S^a$-rotated Bell measurement the updated $a$ key is needed. Measurements must be performed in the pre-established order, because the updated key depends on the result of the previous measurement. Once all measurements are completed and all the keys updated, the client obtains the final key $dk=(a_{\text{final}},b_{\text{final}}), a_{\text{final}}, b_{\text{final}}\in\{0,1\}^n$. Finally the client can decrypt the qubits which are in the state $\ket{\rho_\text{\text{final}}}$ (the final state that the server outputs) using his final updated keys to obtain the plaintext result $\ket{\alpha_{\text{final}}}$: $X^{ a_{\text{final}}}Z^{b_{\text{final}}}\ket{\rho_{\text{final}}}=\ket{\alpha_{\text{final}}}$. 

The whole scheme can be described in five steps: \textbf{Setup}, \textbf{Key Generation}, \textbf{Encryption}, \textbf{Evaluation}, \textbf{Decryption}. 
\begin{enumerate}
	\item \textbf{Setup}:  an EPR source generates $M$ Bell states (as many as $T/T^{\dagger}$ gates are in the circuit) $\{\ket{\Phi_{00}}_{c_i,s_i}, i=1,..,M\}$, where qubits $c_i, i = 1,....,M$ and qubits $s_i, i =
	1,....,M$ are kept by the client and the server respectively.
	\item \textbf{Key Generation}: Generate random bits $a_0$, $b_0$$\in\{0,1\}^n$ and output $sk=(a_0,b_0)$ as the secret key.
	\item \textbf{Encryption}: for any $n$ qubit data $\ket{\alpha}$, client performs
	QOTP encryption with the secret key $sk = (a_0, b_0)$:  $\ket{\alpha} \rightarrow X^{a_0}Z^{b_0} \ket{\alpha}=\ket{\rho_0}$.
	\item  \textbf{Evaluation}: The server applies the quantum gates in order on the $n$ encrypted qubits and updates the key updating function for each qubit according to the key updating algorithm of figure \ref{reglas}. If the gate is a $T/T^{\dagger}$ acting on the $w$-th qubit the server applies a SWAP gate on the $w$-th qubit and one of the entangled qubits $s_i$. After the last gate is applied the server sends all the encrypted qubits, the key-updating functions based on the rules of figure \ref{reglas} and all the ancillary $s_i$ qubits to the client.
    \item  \textbf{Decryption}: the client uses the key updating functions to obtain the corresponding $a$ to the first $S^a$-rotated Bell measurement and measures in the correct basis, obtains $r_a$ and $r_b$ and updates his keys, then repeats the process until there are no more measurements to be made and obtains the final key $dk=(a_{\text{final}},b_{\text{final}})$. Finally the client performs QOTP decryption on the encrypted qubits to obtain the final states: $X^{a_{\text{final}}}Z^{b_{\text{final}}}\ket{\rho_{\text{final}}}=\ket{\alpha_{\text{final}}}$.  
\end{enumerate}

Therefore, this procedure can implement any quantum circuit homomorphically ($\mathcal{F}$-homomorphic), with perfect security since the server can never learn any information about the plaintext or the evaluation keys at any point: the data the server receives is encrypted with QOTP so it is perfectly secure, there are no interactions in the evaluation process so the server can not learn anything about the data there either and once it sends the data to the client for the decryption process it is impossible to obtain any more information.

From all this process the importance of the number of $T$ gates in the circuit becomes clear, because unlike the Clifford gates, they make the complexity of the decryption process grow due to the quantum measurements the client has to perform in succession. This is the reason why the discussed schemes are only suitable for circuits with a polynomial number of $T/T^{\dagger}$ gates. Therefore \textit{T-lineal circuits with reduced circuit complexity (RCC)} are a perfect example of an algorithm that can be evaluated efficiently using Liang's QHE schemes.

\begin{tcolorbox}
	[colback=white]
	\begin{cor}
		\textit{ T-lineal circuits with reduced circuit complexity (RCC)} can be implemented efficiently using Liang's QHE schemes, since their number of $T$ gates grows as $O(n)$, where $n$ is the number of qubits in each register of the circuit and also the number of bits of $s$. 
		\label{cor2}
	\end{cor}
	
\end{tcolorbox}

\section{Conclusions}
\label{sec:conclusions}
In this article the nonrecursive Bernstein-Vazirani algorithm and its recursive counterpart have been reviewed.
The recursive version has historical importance because it showed for the first time a super-polynomial speed-up compared to what classical algorithms could achieve. 

In this paper a certain group of quantum circuits, \textit{ T-lineal circuits with reduced circuit complexity (RCC)}, that solve some particular cases of the recursive Bernstein-Vazirani problem for $k=2$ have been defined and characterized. The main feature of these circuits is simplifying the quantum oracle needed to solve the problem from multi-controlled $CNOT$s to Toffoli gates. This reduces the $T$ gate complexity of the oracle so the number of $T$ gates needed to solve the problem grows linearly with the number of qubits needed to represent the secret key $s$. This property is important in the realm of quantum homomorphic encryption.

Liang's \cite{Liang} quasi-compact Quantum Homomorphic Encryption schemes were reviewed. These schemes had perfect security, $\mathcal{F}$-homomorphism, no interaction between client and server and quasi-compactness. Since the schemes are not compact they do not contradict the no-go result given by Yu \cite{no go result}. The decryption procedure is independent of the size of the evaluated circuit and depends only on the number of $T/T^{\dagger}$ gates in the circuit. The decryption procedure is only efficient for circuits with a polynomial number of $T/T^{\dagger}$ gates. \textit{T-lineal circuits with reduced circuit complexity (RCC)} number of $T/T^{\dagger}$ gates grows linearly as $O(n)$.  
Therefore, they constitute a perfect example of quantum circuits that can be evaluated homomorphically with perfect security and non interaction in an efficient manner.

On the other hand, a circuit that solves the recursive BV problem for $k=2$ that can not be implemented as a \textit{T-lineal circuits with reduced circuit complexity (RCC)} would not be suitable to be implemented using these schemes. This is because as it was shown such a circuit requires an exponential number of $T/T^{\dagger}$ gates, so the decryption process would be inefficient due to the quasi-compactness of the schemes.

Future works in this area of research could be finding if there are some sort of \textit{ T-lineal circuits with reduced circuit complexity (RCC)} for higher orders of the recursion and whether or not these circuits can be evaluated using QHE schemes efficiently. 

Furthermore, more quantum algorithms could be studied regarding their $T$ gates complexity like Grover's algorithm. This way their homomorphic implementation using Liang's schemes could be analysed. It may not be possible to implement them homomorphically in the most general case but particular cases similar to \textit{ T-lineal circuits with reduced circuit complexity (RCC)} where the $T$ gate complexity is low may exist and be useful for certifying instances of quantum computers in the NISQ era where classical simulations can not do otherwise.

\section{Acknowledgements}
We acknowledge the support from the CAM/FEDER Project No.S2018/TCS-4342 (QUITEMAD-CM), Spanish MINECO grants MINECO/FEDER Projects,  PID2021-122547NB-I00 FIS2021, the ``MADQuantum-CM'' project funded by Comunidad de Madrid and by the Recovery, Transformation, and Resilience Plan - Funded by the European Union - NextGenerationEU and Ministry of Economic Affairs Quantum ENIA project. M. A. M.-D. has been partially supported by the U.S.Army Research Office through Grant No. W911NF-14-1-0103. P.F.O. acknowledges support from
a MICINN contract PRE2019-090517 (MICINN/AEI/FSE,UE).

\end{document}